\documentclass[twocolumn,aps,prb,preprintnumbers]{revtex4}
\usepackage{bbm}
\usepackage[latin9]{inputenc}
\usepackage{amsmath}
\usepackage{amssymb}
\usepackage{graphicx}
\usepackage{mathrsfs}
\usepackage{amsfonts}
\usepackage{amsthm}
\usepackage{color}
\usepackage{txfonts}
\providecommand{\U}[1]{\protect\rule{.1in}{.1in}}
\makeatletter
\@ifundefined{textcolor}{}
{
\definecolor{BLACK}{gray}{0}
\definecolor{WHITE}{gray}{1}
\definecolor{RED}{rgb}{1,0,0}
\definecolor{GREEN}{rgb}{0,1,0}
\definecolor{BLUE}{rgb}{0,0,1}
\definecolor{CYAN}{cmyk}{1,0,0,0}
\definecolor{MAGENTA}{cmyk}{0,1,0,0}
\definecolor{YELLOW}{cmyk}{0,0,1,0}
}
\makeatother
\begin{document}
\title{Eavesdropping on spin waves inside the domain-wall nanochannel via three-magnon processes}
\author{Beining Zhang$^{1}$}
\author{Zhenyu Wang$^{1}$}
\author{Yunshan Cao$^{1}$}
\email[Corresponding author: ]{yunshan.cao@uestc.edu.cn}
\author{Peng Yan$^{1}$}
\email[Corresponding author: ]{yan@uestc.edu.cn}
\author{X.R. Wang$^{2,3}$}
\affiliation{$^{1}$School of Electronic Science and Engineering and State Key
Laboratory of Electronic Thin Film and Integrated Devices, University of
Electronic Science and Technology of China, Chengdu 610054, China}
\affiliation{$^{2}$Physics Department, The Hong Kong University of Science and Technology,
 Clear Water Bay, Kowloon, Hong Kong}
\affiliation{$^{3}$HKUST Shenzhen Research Institute, Shenzhen 518057, China}

\begin{abstract}
One recent breakthrough in the field of magnonics is the experimental realization of reconfigurable
spin-wave nanochannels formed by magnetic domain wall with a width of $10-100$ nm [Wagner \emph{et al}., Nat. Nano. \textbf{11}, 432 (2016)]. This remarkable progress enables an energy-efficient spin-wave propagation with a well-defined wave vector along its propagating path inside the wall. In the mentioned experiment, a micro-focus Brillouin light scattering spectroscopy was taken in a line-scans manner to measure the frequency of the bounded spin wave. Due to their localization nature, the confined spin waves can hardly be detected from outside the wall channel, which guarantees the information security to some extent. In this work, we theoretically propose a scheme to detect/eavesdrop on the spin waves inside the domain-wall nanochannel via nonlinear three-magnon processes. We send a spin wave $(\omega_{i},\textbf{k}_{i})$ in one magnetic domain to interact with the bounded mode $(\omega_{b},\textbf{k}_{b})$ in the wall, where $\textbf{k}_{b}$ is parallel with the domain-wall channel defined as the $\hat{z}$ axis. Two kinds of three-magnon processes, i.e., confluence and splitting, are expected to occur. The confluence process is conventional: conservation of energy and momentum parallel with the wall indicates a transmitted wave in the opposite domain with $\omega(\textbf{k})=\omega_{i}+\omega_{b}$ and $(\textbf{k}_{i}+\textbf{k}_{b}-\textbf{k})\cdot\hat{z}=0$, while the momentum perpendicular to the domain wall is not necessary to be conserved due to the non-uniform internal field near the wall. We predict a stimulated three-magnon splitting (or ``magnon laser") effect: the presence of a bound magnon propagating along the domain wall channel assists the splitting of the incident wave into two modes, one is $\omega_{1}=\omega_{b},\textbf{k}_{1}=\textbf{k}_{b}$ identical to the bound mode in the channel, and the other one is $\omega_{2}=\omega_{i}-\omega_{b}$ with $(\textbf{k}_{i}-\textbf{k}_{b}-\textbf{k}_{2})\cdot\hat{z}=0$ propagating in the opposite magnetic domain. Micromagnetic simulations confirm our theoretical analysis. These results demonstrate that one is able to uniquely infer the spectrum of the spin-wave in the domain-wall nanochannel once we know both the injection and the transmitted waves.

\end{abstract}
\maketitle
\section{Introduction}
Spin waves (or magnons) are elementary excitations in ordered magnets. There has been long-term research interest on spin waves ever since they are introduced by Bloch \cite{Bloch} to explain
the celebrated $T^{3/2}$ dependence of spontaneous magnetization on the absolute temperature $T$. In the past few years, intensive investigation on the behaviour of spin waves in nano-structured elements gives birth to an emerging sub-field of condensed matter physics, the magnonics \cite{Grundler,Lenk,Chumak}. The scientific community of magnonics has made huge efforts to achieve concepts to utilize spin waves as data carriers for information processing based on their wave properties \cite{Vogt1,Hillebrands,Vogt2}. On the one hand, it has been proposed that spin waves can efficiently drive the motion of magnetic topological solitons, such as domain walls \cite{Yan} and skyrmions \cite{Iwasaki,Schutte}. On the other hand, spin wave propagation confined in geometrically patterned waveguides has been realized \cite{Urazhdin}. But it lacks the flexibility for controlling the spin-wave propagation path which is required for reprogrammable magnonic devices. From an energy point of view, the dynamic manipulation of spin-waves in two-dimensional structures relies on a continuous application of external forces, e.g. microwaves or spin-polarized currents, and thus demands a high energy consumption. One recent breakthrough is the experimental realization of reconfigurable
spin-wave nanochannels formed by magnetic domain wall with a width of $10-100$ nm \cite{Wagner}. This remarkable progress enables an energy-efficient spin-wave propagating with
a well-defined wave vector along its propagation path inside the wall. Wagner and coworkers used a Brillouin light scattering microscope to locally measure the frequency of the bounded spin waves \cite{Wagner}. Due to their localization nature, the bound spin waves can hardly be detected from outside the wall channel, which guarantees the information security to some extent.

In this work, we propose a non-local scheme to eavesdrop on the spectrum of channelled spin waves via nonlinear three-magnon processes. Three-magnon effects have been known to be important for nonlinear processes in magnetic thin films, since they can give rise to very different output waves \cite{Schultheiss}. For example, in the so-called saturation of ferromagnetic resonance \cite{Suhl}, the uniform mode decays into two modes with a half frequency. Recent spin pumping experiments show that three-magnon processes in magnetic insulators can enhance the interfacial spin-current emission \cite{Demokritov}. Conventional three-magnon processes are triggered by the weak non-local magnetic dipole-dipole interaction in uniform magnetic thin films \cite{Cottam}. There are two different three-magnon-scattering processes: splitting and confluence. Due to conservation of energy the splitting in three magnon-scattering events only occurs, if the pumping frequency is at least twice the frequency of the bottom of the spin wave band. The spin-wave band typically starts at a non-zero frequency and hence, three magnon scattering is prohibited if the pumping frequency is not high enough \cite{Patton}. Exchange coupling and magnetic anisotropy (including both the magneto-crystalline anisotropy and the shape anisotropy due to the local part of the dipolar interaction), on the other hand, are often much stronger than the non-local dipole-dipole interaction in ferromagnet. In homogeneous ferromagnets without external magnetic fields, the lowest-order nonlinear process by these two interactions is the four-magnon scattering \cite{Schultheiss2}. However, three-magnon procsses can occur in magnetic textures such as the skyrmion without the dipolar interaction \cite{Aristov}. Here, we consider three-magnon effect arising in the domain-wall nano-channel (shown in Fig. 1): we input a spin wave $(\omega_{i}, \textbf{k}_{i})$ in one magnetic domain to interact with the mode $(\omega_{b}, \textbf{k}_{b})$ bounded in the domain wall with $\textbf{k}_{b}\parallel \hat{z}$, where $\omega_{i,b}$ and $\textbf{k}_{i,b}$ are the frequency and the wave vector of magnons, respectively. Conservations of both the energy and the momentum parallel with the wall, i.e., $\omega(\textbf{k})=\omega_{i}+\omega_{b}$ and $(\textbf{k}_{i}+\textbf{k}_{b}-\textbf{k})\cdot\hat{z}=0$, enable us to uniquely determine the spectrum $(\omega, \textbf{k})$ of the three-magnon confluence. We note that the momentum perpendicular to the domain wall is not necessary to be conserved due to the non-uniform internal field near the wall. On the other hand, when the frequency of incident magnons goes beyond a threshold value, the three-magnon splitting emerges as well, i.e., $\omega_{i}\rightarrow\omega_{1}+\omega_{2}$ and $\textbf{k}_{i}\rightarrow\textbf{k}_{1}+\textbf{k}_{2}$. In general, the mentioned two conservation laws are insufficient to uniquely determine the splitting spectrum. However, the presence of the bounded magnon stimulates a ``magnon laser" effect which makes one of the two split modes to be identical to the bound mode, i.e., $\omega_{1}=\omega_{b}$ and $\textbf{k}_{1}=\textbf{k}_{b}$. The other mode $(\omega_{2}, \textbf{k}_{2})$ then can be uniquely determined from the energy-momentum conservations, i.e., $\omega_{2}=\omega_{i}-\omega_{b}$ and $(\textbf{k}_{i}-\textbf{k}_{b}-\textbf{k}_{2})\cdot\hat{z}=0$. These results demonstrate that, with the help of the information of injection wave in one magnetic domain and the emerging modes in another domain, we are able to uniquely infer the spectrum of the spin-wave in the nanochannel formed by the domain wall. Micromagnetic simulations are implemented to verify our theoretical results.

This paper is organized as follows. In Sec. II, the theoretical consideration based on the Landau-Lifshitz phenomenology is presented. Spectrum of linear spin waves is given on top of a two-dimensional domain wall structure. Three-magnon processes arising inside the wall channel are analyzed as well. Section III gives the results of micromagnetic simulations to verify the theoretical predictions. Conclusions are drawn in Sec. IV. Magnon-magnon interaction Hamiltonian in inhomogeneous magnetization textures is derived in the Appendix.

\section{Theoretical considerations}

We start with the Hamiltonian
\begin{equation}\label{Hamiltonian}
 H=\int d\textbf{r} \bigg[\frac{A}{M^{2}_{s}} {(\nabla \textbf{M})}^{2}-\frac{D}{M^{2}_{s}} (\textbf{M}\cdot \textbf{n})^2\bigg],
\end{equation}\begin{figure}[ptbh]
\begin{centering}
\includegraphics[width=0.52\textwidth]{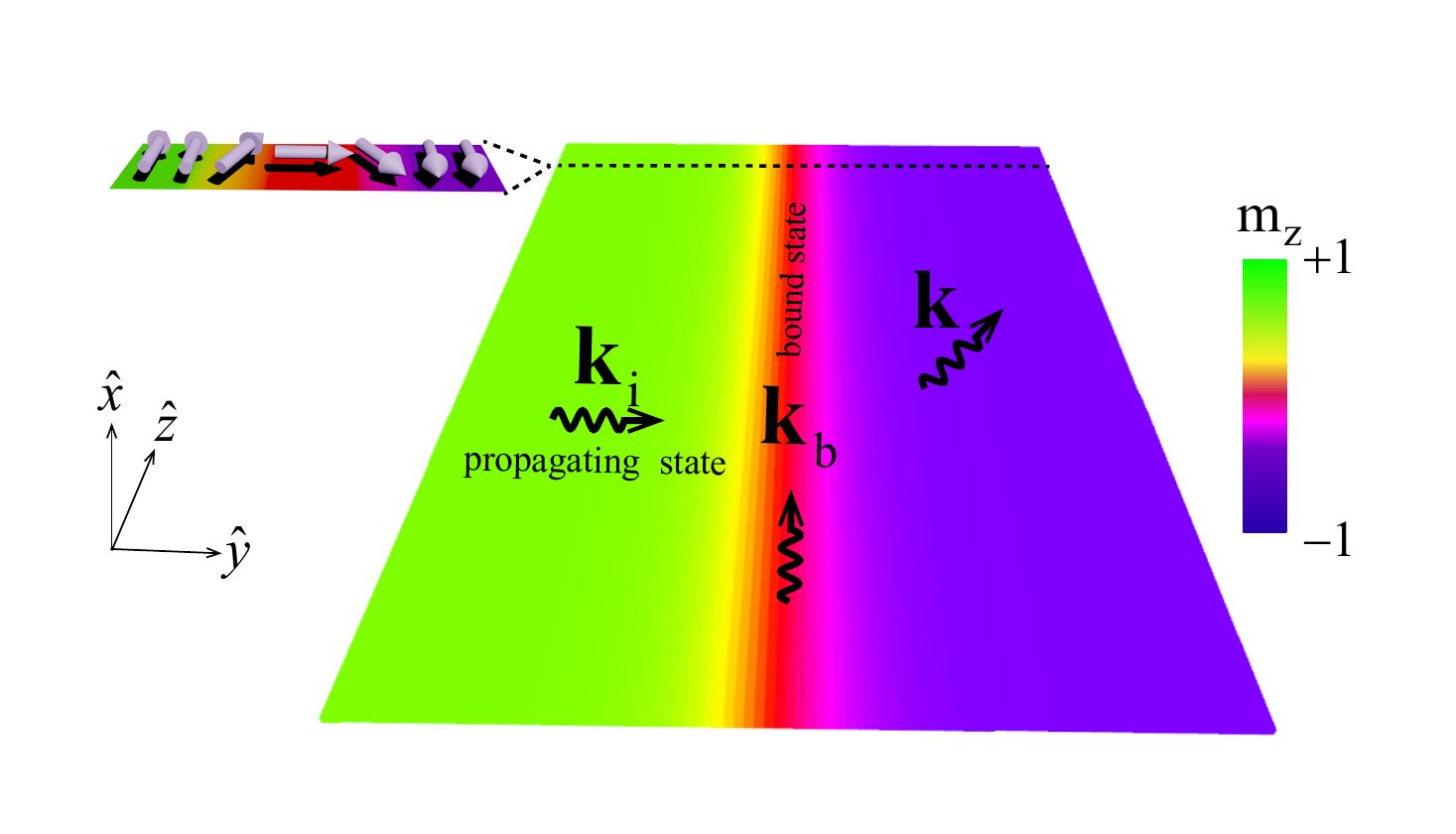}
\par\end{centering}
\caption{Schematic plot of three-magnon processes arising inside a N\'{e}el domain wall in an extended thin ferromagnetic film. $\textbf{k}_{b}$ is the wave vector of the bounded spin-wave propagating along the domain-wall channel, while $\textbf{k}_{i}$ is the wave vector of bulk spin-wave in the magnetic domain. Three-magnon confluence generates an emerging mode with wave vector $\textbf{k}$. The domain wall structure is zoomed in at the top left coner. Three-magnon splitting process is not shown.}
\end{figure}in two spatial dimensions. Here $\textbf{M}=M_{s}\textbf{m}$ is the magnetization with the saturated value $M_{s}$ and the direction $\textbf{m}$, $A$ is the exchange constant, $D$ is the anisotropy constant, and $\textbf{n}$ is the unit vector along the anisotropy axis (the $z$ axis). In the theoretical analysis, the magnetic dipole-dipole interaction is ignored for simplicity, but it can be included in numerical calculations in the next section. Based on this energy functional, we consider a magnetic thin film with two magnetic domains, whose magnetizations point in opposite directions separated by a N\'{e}el domain wall, as shown in Fig. 1. The film is in the $y$-$z$ plane and the magnetization in the left/right domain is along the $\pm \hat{z}$ direction, i.e., $\textbf{m}(y=\pm\infty)=\mp \hat{z}$, respectively. The nanochannel formed by the domain wall is along the $\hat{z}$ direction as well. Minimizing the energy functional with the mentioned boundary condition gives rise to the so-called Walker solution \cite{Walker}
\begin{equation}\label{Walker}
  m_{0,x}=0, m_{0,y}=\frac{1}{\cosh\frac{y-Y}{w}}, \text{ and }m_{0,z}=-\tanh\frac{y-Y}{w},
\end{equation}
describing the spatial distribution of static domain wall magnetization $\textbf{m}_{0}$. Here $Y$ is the position of domain wall center and $w=\sqrt{A/D}$ is the domain wall width. Spatiotemporal evolution of dynamic magnetization is governed by the classical Landau-Lifshitz-Gilbert (LLG) equation,
\begin{equation}\label{LLG}
  \frac{\partial\textbf{m}(\textbf{r},t)}{\partial t}=-\gamma \textbf{m}\times \textbf{H}_{\text{eff}}+\alpha\textbf{m}\times\frac{\partial\textbf{m}}{\partial t},
\end{equation}
where $\gamma$ is the gyromagnetic ratio, $\alpha\ll1$ is the dimensionless Gilbert damping constant, and $\textbf{H}_{\text{eff}}=-\mu^{-1}_{0}\delta H/\delta \textbf{M}$ is the effective magnetic field with vacuum permeability $\mu_{0}$. We first derive the linear spin-wave spectrum on top of the static domain wall. To this end, we assume a small fluctuation of $\textbf{m}$ around $\textbf{m}_{0}$, and express $\textbf{m}$ in local spherical coordinates $\textbf{e}_{r},\textbf{e}_{\theta}$, and $\textbf{e}_{\phi}$ as $\textbf{m}(\textbf{r},t)\approx\textbf{m}_{0}+m_{\theta}(\textbf{r},t)\textbf{e}_{\theta}+m_{\phi}(\textbf{r},t)\textbf{e}_{\phi}$ with $\textbf{e}_{r}\equiv\textbf{m}_{0}$ and $|m_{\theta,\phi}|\ll 1$. By defining a wave function $\psi(\textbf{r},t)=m_{\theta}(\textbf{r},t)-im_{\phi}(\textbf{r},t)$ and neglecting the Gilbert damping, the LLG equation (\ref{LLG}) can be linearized and recast into a Schr\"{o}dinger-like equation \cite{Yan}
\begin{equation}\label{Schrodinger}
  i\hbar\frac{\partial\psi}{\partial \tau}=\bigg[\frac{\hat{\textbf{p}}^{2}}{2m^{*}}+V(\textbf{r})\bigg]\psi,
\end{equation}
with $\tau=\gamma t/\mu_{0}M_{s}$, the effective mass $m^{*}=\hbar/4A$, the momentum operator $\hat{\textbf{p}}=-i\hbar\nabla$, and the reflectionless potential well $V(\textbf{r})=2D\hbar[1-2\cosh^{-2}(\frac{y-Y}{w})]$ with $\hbar$ the reduced Planck constant. One should note that $\hbar$ can be completely eliminated by dividing it on both sides of Eq. (\ref{Schrodinger}) which thus is not a true quantum-mechanical Schr\"{o}dinger equation. However, its solutions can indicate interesting physics. Equation (\ref{Schrodinger}) allows two types of solutions. One is the scattering spin-wave state with \cite{Bayer}
\begin{equation}\label{Propagating}
\begin{aligned}
  \psi_{\textbf{k},i}(\textbf{r},t)=&\frac{\tanh(\frac{y-Y}{w})-ik_{y}w}{\sqrt{1+w^{2}k^{2}_{y}}}e^{i(k_{y}y+k_{z}z)-i\omega_{i} t},\\
  \omega_{i}=&\frac{2\gamma}{\mu_{0}M_{s}}[D+A(k^{2}_{y}+k^{2}_{z})],
\end{aligned}
\end{equation}
while the other one corresponds to a spin wave localized near the domain wall
\begin{equation}\label{Localized}
  \begin{aligned}
  \psi_{\textbf{k},b}(\textbf{r},t)=&\frac{1}{\sqrt{2}}\text{sech}(\frac{y-Y}{w})e^{ik_{z}z-i\omega_{b} t},\\
  \omega_{b}=&\frac{2\gamma}{\mu_{0}M_{s}} Ak^{2}_{z}.
\end{aligned}
\end{equation}
The set of functions $\psi_{\textbf{k}}(\textbf{r},t)$ is complete and orthonormal. Important physics associated with both the scattering (\ref{Propagating}) and the bound states (\ref{Localized}) have been exploited in recent literatures. For instance, the reflectionless property of the scattering states leads to the so-called all-magnonic spin transfer torque \cite{Yan}, while the realization of spin-wave propagation along the domain-wall nanochannel numerically \cite{Jiang,Kim} and experimentally \cite{Wagner} relies on the localized nature of the bound states. However, the interplay between these two modes is yet to be addressed. We present the rigorous derivation of the magnon-magnon interaction Hamiltonian in the Appendix, and here focus on the physics of three-magnon effect which is the lowest-order nonlinear process in inhomogeneous magnetization textures, even without the magnetic dipolar interaction. As shown in Fig. 1, we input a propagating wave $(\omega_{i},\textbf{k}_{i})$ in the left magnetic domain, to interact with the bounded mode $(\omega_{b},\textbf{k}_{b})$. In general, both the confluence and the splitting events can occur in the three-magnon processes, as shown in Fig. 2.
\begin{figure}[ptbh]
\begin{centering}
\includegraphics[width=0.35\textwidth]{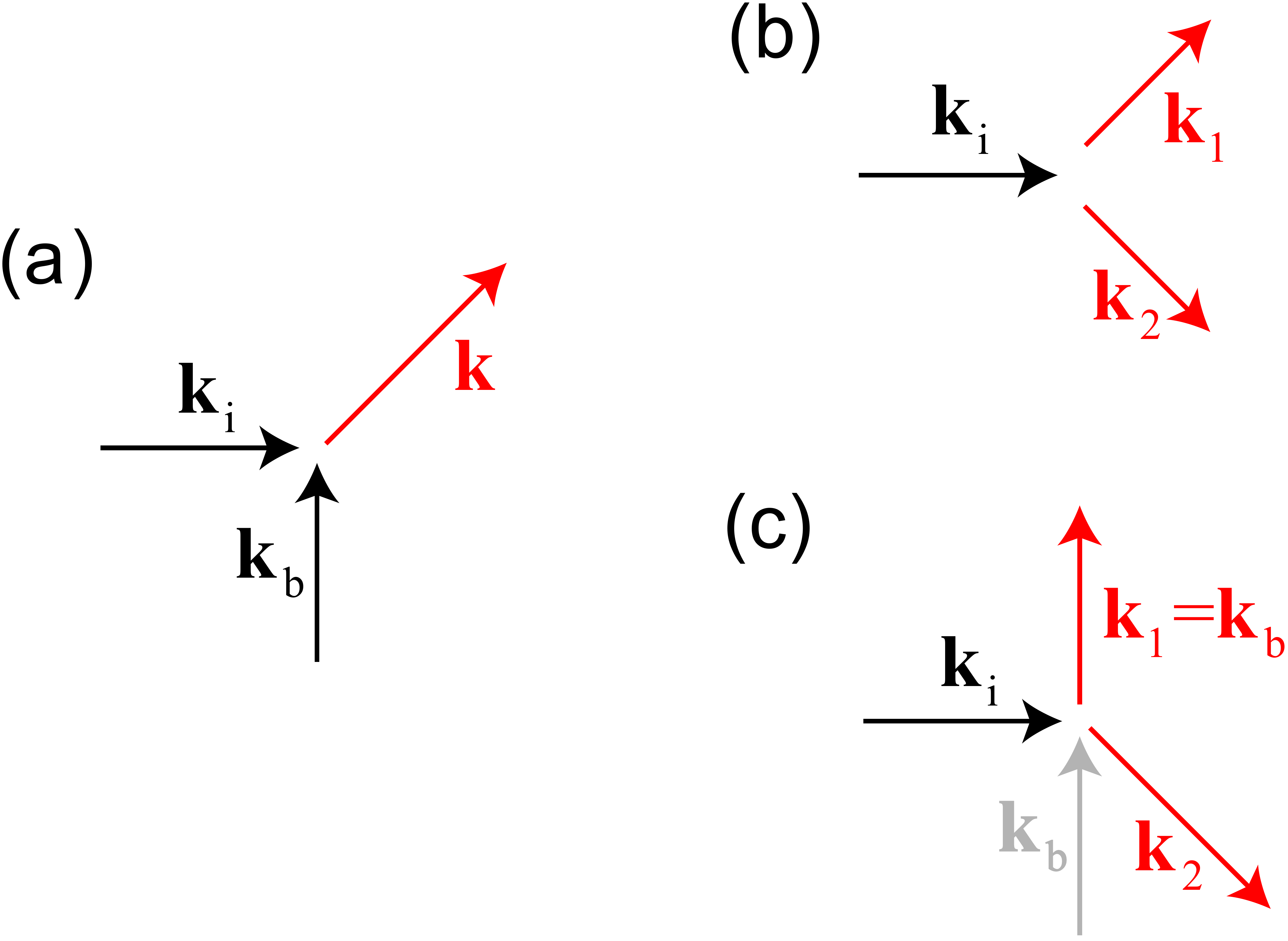}
\par\end{centering}
\caption{Schematic picture of nonlinear three-magnon processes. (a) Three-magnon confluence of $\textbf{k}_{i}$ and $\textbf{k}_{b}$ into $\textbf{k}$. (b) Spontaneous three-magnon splitting of $\textbf{k}_{i}$ into two random modes $\textbf{k}_{1}$ and $\textbf{k}_{2}$. (c) Stimulated three-magnon splitting of $\textbf{k}_{i}$ into two modes $\textbf{k}_{1}=\textbf{k}_{b}$ and $\textbf{k}_{2}$, assisted by a localized magnon $\textbf{k}_{b}$ (grey arrow).}
\end{figure}

\subsection{Three-magnon confluence}
We first consider the three-magnon confluence event shown in Fig. 2(a). In this process, the energy is conserved (under the assumption of negligibly small dissipation due to the Gilbert damping) while the particle number is not. Translational invariance along $z$ direction guarantees the conservation of momentum parallel with the domain wall. We thus have
\begin{equation}\label{Confluence}
  \begin{array}{c}
  \omega_{\textbf{k}}=\omega_{i}+\omega_{b},\\
  (\textbf{k}-\textbf{k}_{i}-\textbf{k}_{b})\cdot \hat{z}=0.
  \end{array}
\end{equation}
Considering propagating wave with an arbitrary incident angle $\beta$, i.e., $\textbf{k}_{i}=|\textbf{k}_{i}|(\cos\beta \hat{y}+\sin\beta \hat{z})$, we obtain the solution of the confluence spectrum $\textbf{k}=(|\textbf{k}_{i}|\cos\beta+q)\hat{y}+(|\textbf{k}_{i}|\sin\beta+k_{b})\hat{z}$,
where the parameter $q$ measures the momentum mismatch perpendicular to the wall and satisfies the following equation
\begin{equation}\label{qq}
  q^{2}+2(|\textbf{k}_{i}|\cos\beta) q+2|\textbf{k}_{i}|\sin\beta k_{b}=0.
\end{equation}
For a normal incident, i.e., $\beta=0$, we obtain $q=0$ and $q=-2|\textbf{k}_{i}|$ which corresponds to a forward confluence
\begin{equation}\label{Forward}
  \textbf{k}=\textbf{k}_{i}+\textbf{k}_{b},
\end{equation}
and a backward one
\begin{equation}\label{Backward}
  \textbf{k}=-\textbf{k}_{i}+\textbf{k}_{b},
\end{equation}
respectively. The intensity of the three-magnon confluence process is given by $I_{\text{con}}\propto n_{\textbf{k}_{i}}n_{\textbf{k}_{b}}$ \cite{Akhiezer}, with $n_{\textbf{k}_{i}}$ and $n_{\textbf{k}_{b}}$ the numbers of magnons in the initial states. In the classical region, $n_{\textbf{k}_{i,b}}\gg1$.

\subsection{Three-magnon splitting: random and stimulated}
Figure 2(b) shows a general three-magnon splitting process of the incident wave $(\omega_{i},\textbf{k}_{i})$. In this process, the energy-momentum conservation gives rise to
\begin{equation}\label{ConservationSp}
\begin{array}{c}
  \omega_{1}+\omega_{2}=\omega_{i},\\
  (\textbf{k}_{1}+\textbf{k}_{2}-\textbf{k}_{i})\cdot \hat{z}=0,
\end{array}
\end{equation}
with the intensity given by $I_{\text{spl}}\propto n_{\textbf{k}_{i}}$ \cite{Akhiezer} with  $n_{\textbf{k}_{i}}$ the magnon number in the initial state, which is much smaller than the intensity of the confluence process. Furthermore, the solution of Eqs. (\ref{ConservationSp}) is obviously not unique. We thus call this process as a random (or spontaneous) three-magnon splitting. However, the presence of a bound magnon propagating along the wall can trigger a stimulated three-magnon splitting, making one of the two split modes to be identical to the localized mode, i.e.,
\begin{equation}\label{StimulatedM}
  \textbf{k}_{1}=\textbf{k}_{b} \text{ and } \omega_{1}=\omega_{b}.
\end{equation}
In analogy to the stimulated emission of electromagnetic radiation, we call this process a ``magnon laser" effect. The spectrum intensity $I_{\text{spl}}$ then will be significantly enhanced by the very presence of the stimulating modes, and will increase with the increasing $n_{\textbf{k}_{b}}$ (see numerical evidences in Fig. 9 below). While a microscopic perturbative calculation is not the scope of the present work, we note that the concept of stimulated emission was first introduced by Einstein in his seminal derivation of the blackbody spectrum \cite{Einstein}. It is sometimes regarded as pure quantum-mechanical effect. However, it has been pointed out that the stimulated emission arises also in purely classical nonlinear systems by Gaponov \cite{Gaponov}, Fain \cite{Fain1}, and Fain and Milonni \cite{Fain2}. The stimulated emission is understood as a constructive interference between the incident wave and the wave scattered \cite{Cray}. In our stimulated three-magnon splittings, the localized magnon acts as the incident wave, while the impinging magnon corresponds to the wave scattered. The other mode then can be uniquely determined by $\textbf{k}_{2}=(|\textbf{k}_{i}|\cos\beta+q)\hat{y}+(|\textbf{k}_{i}|\sin\beta-k_{b})\hat{z}$,
with parameter $q$ the solution of the following equation
\begin{equation}\label{qqq}
  q^{2}+(2|\textbf{k}_{i}|\cos\beta)q+2k^{2}_{b}-2|\textbf{k}_{i}|\sin\beta k_{b}=0.
\end{equation}
We are again interested in the normal-incident case. Then the above equation is reduced to $q^{2}+2|\textbf{k}_{i}|q+2k^{2}_{b}=0$ which allows real solutions $q=-|\textbf{k}_{i}|\pm\sqrt{|\textbf{k}_{i}|^{2}-2k^{2}_{b}}$ only when
\begin{equation}\label{Criterion}
  |\textbf{k}_{i}|\geqslant\sqrt{2}k_{b}.
\end{equation}
We thus obtain
\begin{equation}\label{SolutionStSp2}
  \textbf{k}_{2}=\pm\sqrt{|\textbf{k}_{i}|^{2}-2k^{2}_{b}}\hat{y}-\textbf{k}_{b},
\end{equation}
corresponding to the forward (``$+$" sign) and the backward (``$-$" sign) splitting solutions. We focus on the forward one in this work.
\section{Numerical results}

\begin{figure}[ptbh]
\begin{centering}
\includegraphics[width=0.45\textwidth]{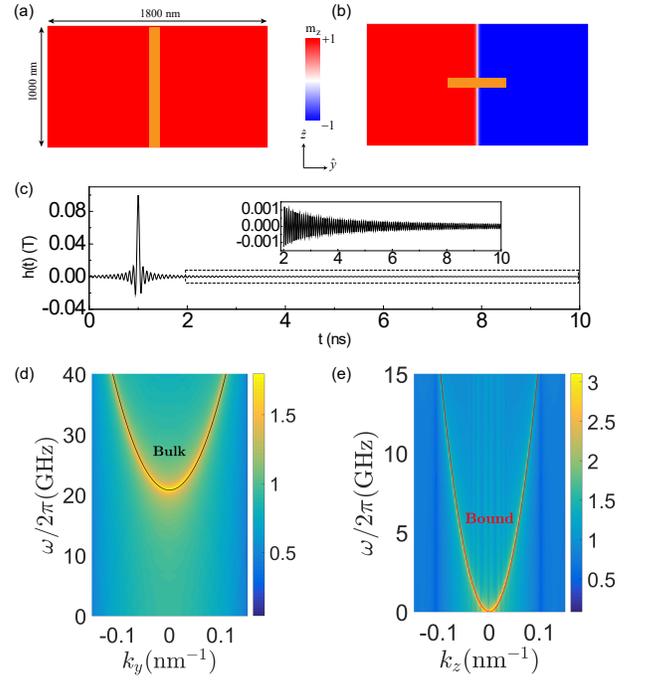}
\par\end{centering}
\caption{Geometry in numerically calculating the linear spin-wave spectrum. (a) Single magnetic domain. (b) Magnetic thin film with a domain wall. (c) Time dependence of the sinc-function field $\textbf{h}(t)$. (d) FFT spectrum over the lattices along $z=500$ nm in (a). (e) FFT spectrum along the domain-wall channel $y=Y=900$ nm in (b). In (d) and (e), the solid curves are the analytical formula (\ref{Propagating}) and (\ref{Localized}), respectively, without any fitting parameter. Microwave driving fields are located in the regions of orange color in (a) and (b).}
\end{figure}

To verify our theoretical analysis, we solve numerically the full LLG equation (\ref{LLG}) using the micromagnetic simulation codes MuMax3 \cite{Mumax}. We used magnetic parameters of Co with an exchange constant $A=4\times10^{-11}$ J m$^{-1}$, a uniaxial anisotropy $D=5.2\times10^{5}$ J m$^{-3}$, a saturated magnetization $M_{s}=1.4\times10^{6}$ A m$^{-1}$, a gyromagnetic ratio $\gamma=2.21\times10^{5}$ rad s$^{-1}$ m A$^{-1}$, and a Gilbert damping constant $\alpha=0.02$. The geometry is illustrated in Fig. 3. The magnetic thin film lies in the $y$-$z$ plane, with length 1800 nm, width 1000 nm, and thickness 2 nm, which was discretized using $900\times500\times1$ finite difference cells. Figure 3(a) and (b) show the film without and with a N\'{e}el domain wall, respectively. We first simulate the linear spin-wave spectrum. To this end, we apply a microwave driving field with the sinc-function $\textbf{h}(t)=h_{0}\sin[\omega_{H}(t-t_{0})]/[\omega_{H}(t-t_{0})]\hat{x}$ for 10 ns with $h_{0}=0.1$ T, $\omega_{H}/2\pi=$ 80 GHz and $t_{0}=1$ ns, over the regions of orange color with volumes $30\times1000\times2$ nm$^{3}$ and $30\times400\times2$ nm$^{3}$ shown in Figs. 3(a) and (b), respectively. Figure 3(c) shows the time dependence of the excitation field. The spatiotemporal oscillation of the out-of-plane magnetization component $M_{x}$ is analyzed over the lattices along $z=500$ nm in Fig. 3(a), and over the lattices in the domain-wall center, i.e., $y=Y=900$ nm, in Fig. 3(b). The corresponding fast Fourier transformation (FFT) spectrum are plotted in Figs. 3(d) and (e), respectively. The frequency resolution of the FFT is 0.1 GHz. Numerical results agree excellently with the analytical formula Eqs. (\ref{Propagating}) and (\ref{Localized}) [solid curves shown in Figs. 3(d) and (e)]. In Fig. 3(e), we did not plot the spectrum very close to the gap $\gamma D/(\pi\mu_{0}M_{s})=20.79$ GHz, because in higher frequencies the confinement of spin waves becomes worse and one cannot clearly identify the localized mode from the FFT. We therefore only show the frequency up to 15 GHz.

\begin{figure}[ptbh]
\begin{centering}
\includegraphics[width=0.45\textwidth]{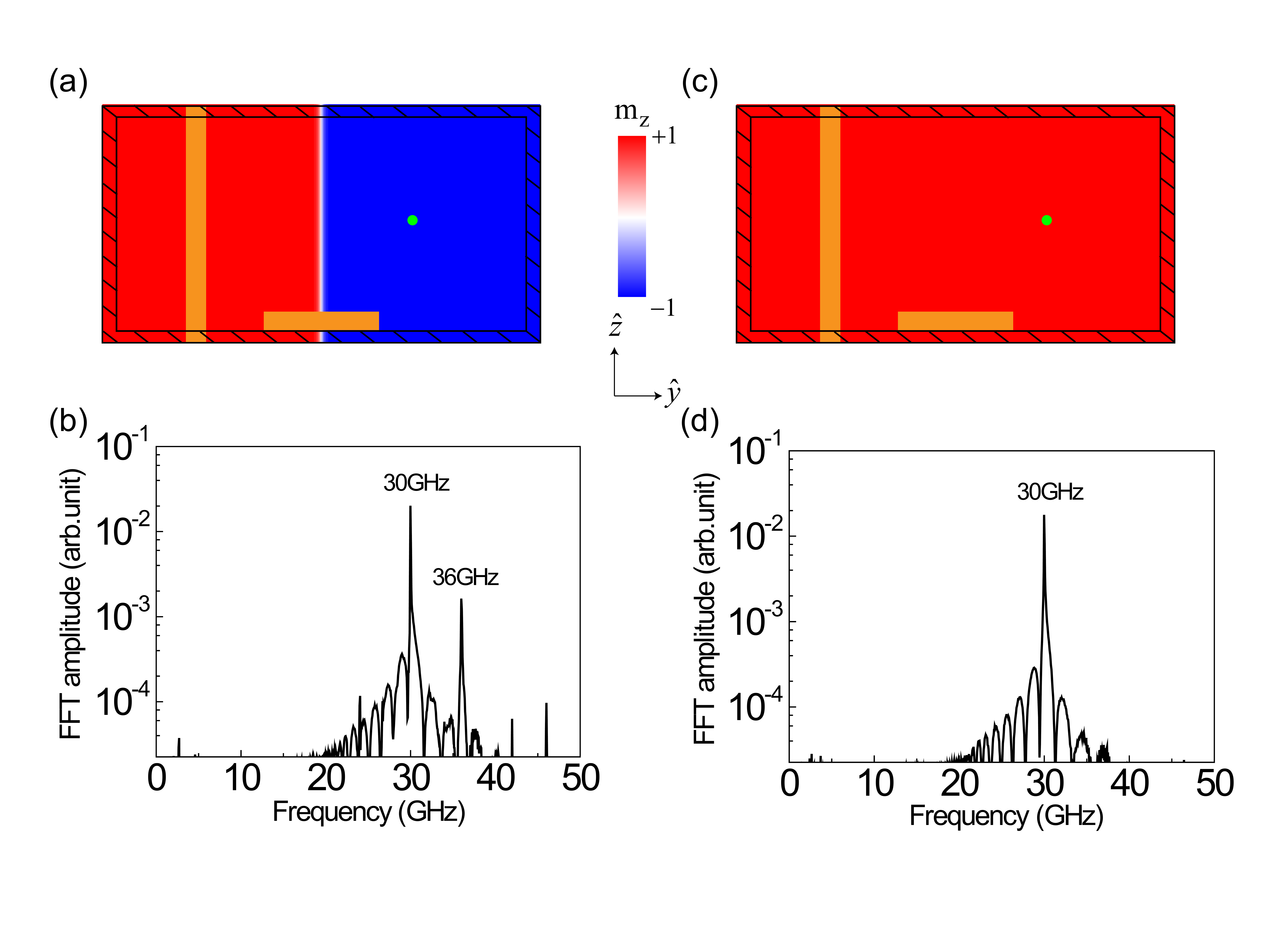}
\par\end{centering}
\caption{Setup to simulate the nonlinear three-magnon processes. (a) Magnetic thin film with a domain wall. (b) FFT spectrum at a single lattice cell [green dot in (a)]. (c) Single magnetic domain. (d) FFT of a single lattice cell [gree dot in (c)]. External microwave fields are located in the regions of orange color in (a) and (c). Absorbing boundary conditions are adopted in the dashed area near the film edges.}
\end{figure}
Then, we simulate the interaction between the propagating and the localized spin waves. We focus on the normal incident case. To this end, we put two sinusoidal monochromatic microwave sources simultaneously over the magnetic film [orange regions shown in Fig. 4(a)]: one source is $\textbf{h}_{i}(t)=h_{i}\sin(\omega_{i}t)\hat{x}$ put in the left domain and the other one is $\textbf{h}_{b}(t)=h_{b}\sin(\omega_{b}t)\hat{x}$ located at the bottom of the film across the domain wall, where $\omega_{i(b)}$ should be well above (below) the band gap of bulk spin waves. We set $h_{i}=h_{b}=h_{0}$ unless otherwise stated. A gradient in the damping constant is utilized at the film edges [dashed area shown in Fig. 4(a)] to avoid the artificial spin-wave reflections by the boundaries \cite{ABC}. We consider $\omega_{i}/2\pi=$30 GHz and $\omega_{b}/2\pi=$6 GHz (much lower than the band gap of bulk spin-waves). The excited magnons carry wave vectors $\textbf{k}_{i}=0.078\hat{y}$ and $\textbf{k}_{b}=0.06\hat{z}$ in unit of nm$^{-1}$, respectively. FFT spectrum analysis at a single cell [the green dot in Fig. 4(a)] shows two peaks at 30 GHz and 36 GHz, respectively, as plotted in Fig. 4(b). The main peak of 30 GHz is from the propagating spin-wave generated by the microwave source in the left domain. While we infer that the relatively weak peak at 36 GHz is due to the three-magnon confluence process, because ``$36=30+6$", there, however, still exists a loophole in this argument: There are two microwave sources $\textbf{h}_{i}(t)$ and $\textbf{h}_{b}(t)$ acting on the ferromagnet, so it is possible that the output spin wave with the sum-frequency could be simply due to the combined driving of the microwave fields on the magnetic moment, rather than the interaction between the propagating wave and the localized wave moving in the domain wall. To close this loophole, we consider a single domain setup shown in Fig. 4(c), without changing rest conditions. The FFT analysis at the same lattice cell [the green dot in Fig. 4(c)] clearly shows the disappearance of the 36 GHz mode without the domain wall, as plotted in Fig. 4(d). This concludes that the emerging high-frequency mode must come from the interaction between the propagating wave and the wave bounded in the domain-wall channel.

\begin{figure}[ptbh]
\begin{centering}
\includegraphics[width=0.5\textwidth]{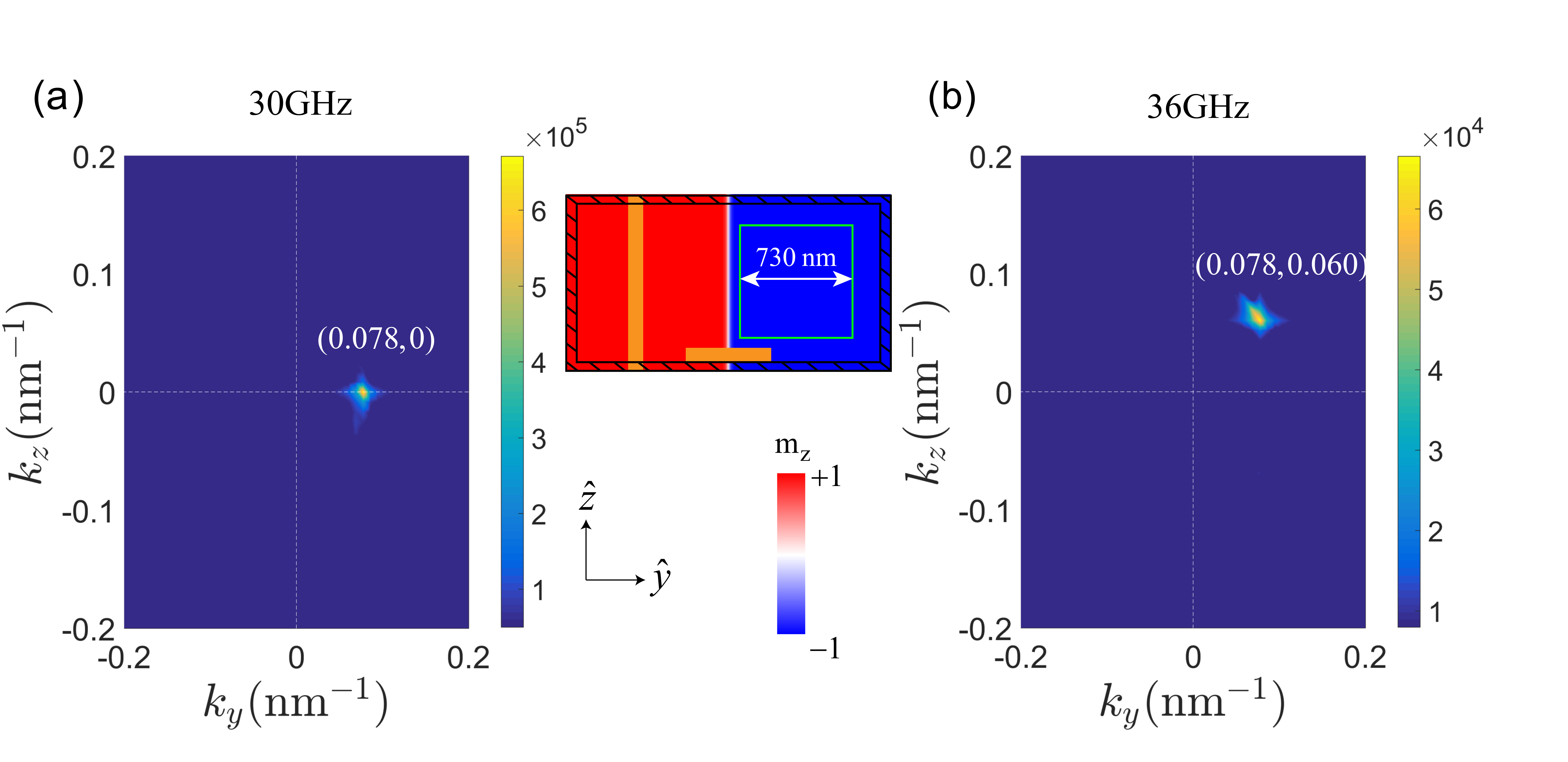}
\par\end{centering}
\caption{Spatial FFT spectrum for the two peaks (a) 30 GHz and (b) 36 GHz, observed in Fig. 4(b) where the incident magnon frequency is $\omega_{i}/2\pi=30$ GHz and the bound magnon frequency is $\omega_{b}/2\pi=6$ GHz. The FFT analysis are implemented over the region inside the green square with the side length 730 nm.}
\end{figure}

To provide a direct evidence, we implement spatial FFT spectrum analysis for the two frequency peaks over the region inside the green square with the side length 730 nm plotted in Fig. 5. The spatial resolution of the FFT spectrum is 0.009 nm$^{-1}$. FFT results are shown in Figs. 5(a) and (b). The magnon wave vectors at 30 GHz and 36 GHz are $\textbf{k}_{i}=0.078\hat{y}$ and $\textbf{k}=0.078\hat{y}+0.06\hat{z}$ in unit of nm$^{-1}$, respectively. These numbers excellently agree with the forward three-magnon confluence formula Eq. (\ref{Forward}).

As we stated earlier, one advantage to use the domain-wall channelled spin-wave is its economic energy consumption. To demonstrate this, we compute the driving powers to excite both the impinging and the bound waves. The instantaneous power is written as $P_{i,b}(t)=-\int_{V_{i,b}}\textbf{M}\cdot \dot{\textbf{h}}_{i,b}(t)d\textbf{r}$ with $V_{i(b)}$ the volume of the ferromagnet covered by microwave sources (as shown in Fig. 5) to generate the impinging (bound) waves. Its time average as functions of both the field amplitude and the frequency is plotted in Fig. 6. In the calculations, we open both microwave sources to mimic the situation of detection. We find that the mean power to excite bound spin waves is two orders of magnitude smaller than that needed to generate the impinging waves. As an example, $P_{i}=86$ nW has to be consumed to excite the impinging wave, while $P_{b}$ is 0.89 nW only to generate the bound wave investigated in Fig. 5.

\begin{figure}[ptbh]
\begin{centering}
\includegraphics[width=0.48\textwidth]{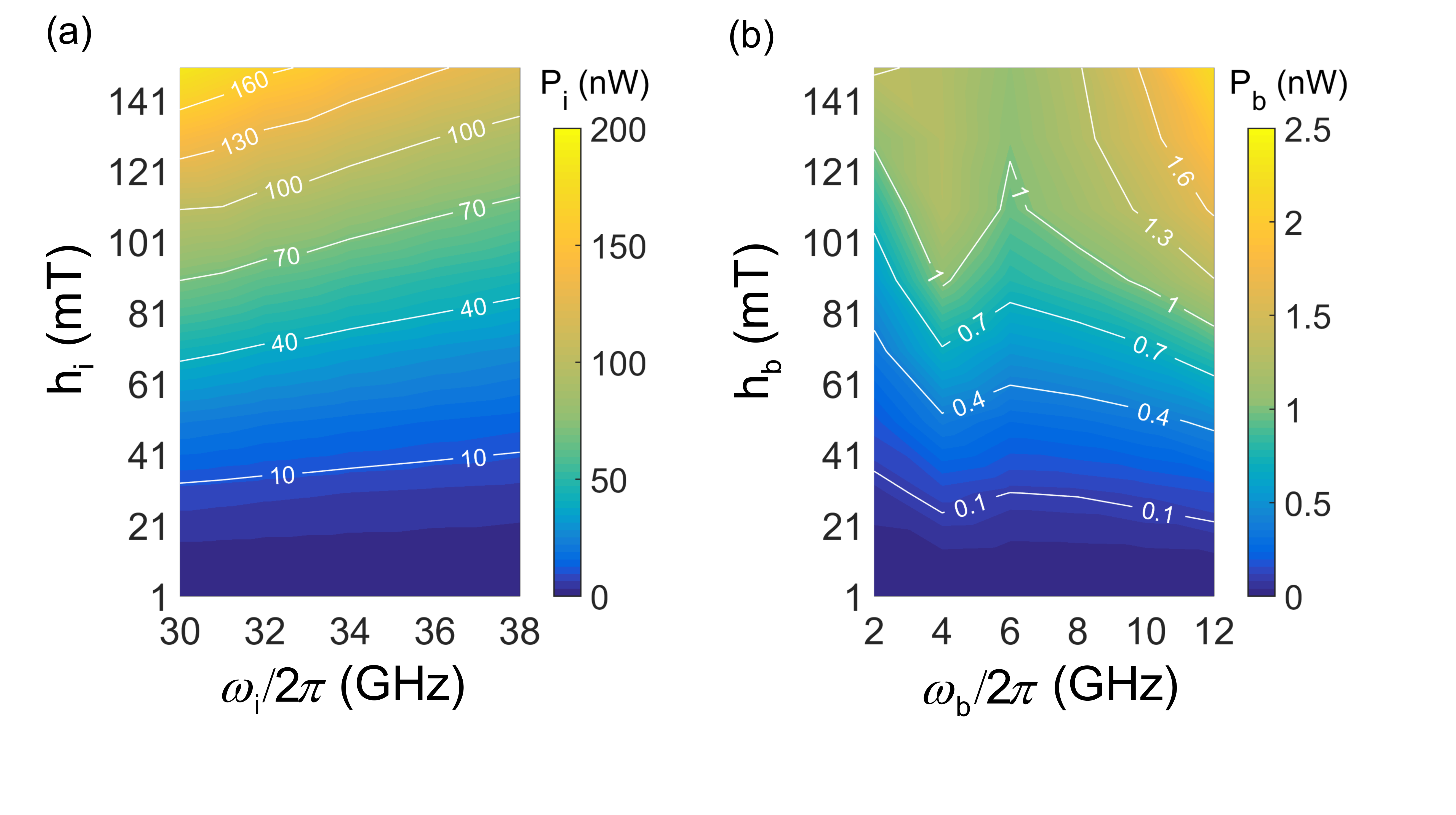}
\par\end{centering}
\caption{Dependence on the field amplitude and the frequency of the driving power (a) $P_{i}$  to generate the impinging spin waves under $h_{b}=100$ mT and $\omega_{b}/2\pi=6$ GHz, and (b) $P_{b}$ to excite the bounded spin waves with $h_{i}=100$ mT and $\omega_{i}/2\pi=30$ GHz.}
\end{figure}

\begin{figure}[ptbh]
\begin{centering}
\includegraphics[width=0.48\textwidth]{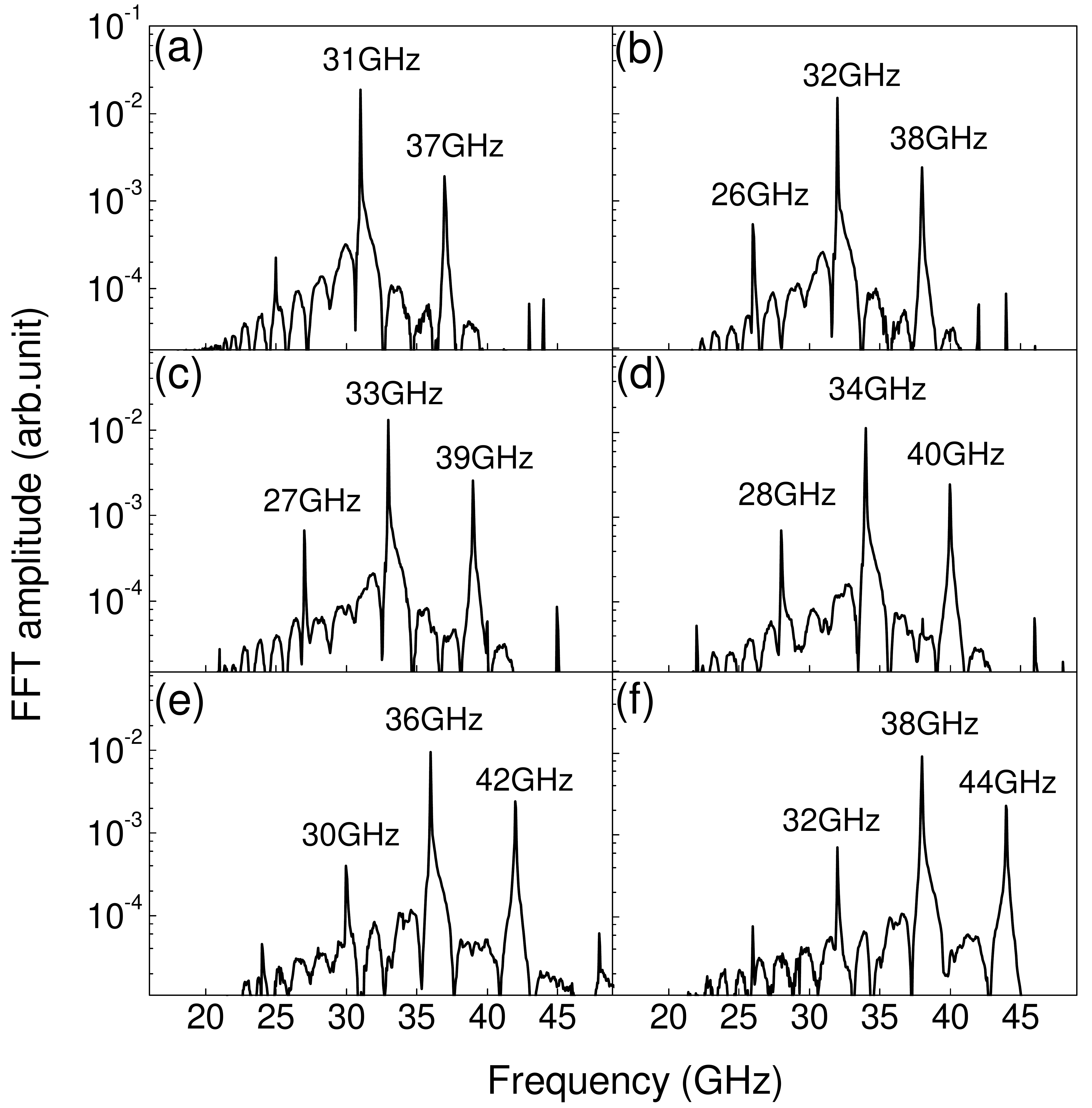}
\par\end{centering}
\caption{Temporal FFT spectrum analysis for six different incident wave frequencies (a) $\omega_{i}/2\pi=31$ GHz. (b) $\omega_{i}/2\pi=32$ GHz. (c) $\omega_{i}/2\pi=33$ GHz. (d) $\omega_{i}/2\pi=34$ GHz. (e) $\omega_{i}/2\pi=36$ GHz. (f) $\omega_{i}/2\pi=38$ GHz.}
\end{figure}

Up to now, signals associated with the three-magnon splitting process, however, did not appear yet. According to the criterion (\ref{Criterion}), we expect the emergence of stimulated splittings when the incident frequency $\omega_{i}/2\pi$ is higher than 32.5 GHz under a fixed $\omega_{b}/2\pi=$6 GHz. We therefore systematically increase the frequency of the incident wave from 30 GHz to 38 GHz in the simulations. Numerical results are shown in Figs. 7(a)-(f), from which we observe a new peak (with FFT amplitude larger than $3\times10^{-4}$) emerging in the low frequency side when the incident wave frequency is no less than 32 GHz, besides the main peak due to the incident wave and the peak in the high frequency side because of the three-magnon confluence process discussed above. The threshold frequency obtained numerically is consistent with the theoretical prediction (\ref{Criterion}) with a discrepancy less than 1.5\%. The distance from the new peak to the main peak is again exactly the frequency of the bound mode. We have interpreted this in terms of a stimulated three-magnon splitting. We would like to remark that the small peak at $25$ GHz in Fig. 7(a) was excluded from the splitting process, because the FFT amplitude is too weak on the one hand and the corresponding momentum violates the splitting solution (\ref{SolutionStSp2}) on the other hand (see below). We attribute it to higher order spin wave processes. To provide more evidences, we do the spatial FFT spectrum analysis for the case $\omega_{i}/2\pi=34$ GHz. Numerical results are shown in Figs. 8(a)-(c). The obtained wave vector at 28 GHz perfectly fits Eq. (\ref{SolutionStSp2}). Spatial FFT analysis on other frequencies $\omega_{i}/2\pi=32,33,36$ and $38$ GHz (except $\omega_{i}/2\pi=31$ GHz) supports the same conclusion.

\begin{figure}[ptbh]
\begin{centering}
\includegraphics[width=0.48\textwidth]{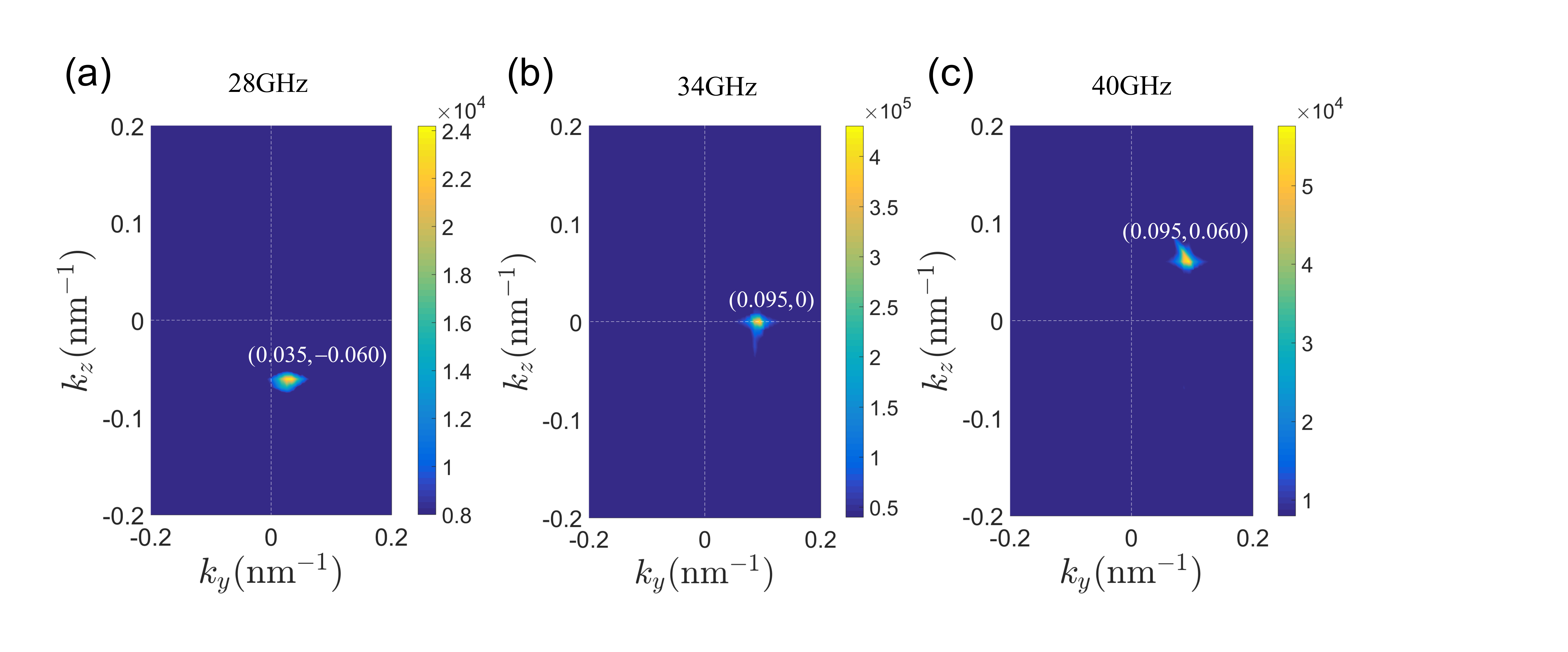}
\par\end{centering}
\caption{Spatial FFT spectrum analysis for the three peaks at (a) 28 GHz, (b) 34 GHz, and (c) 40 GHz, observed in Fig. 7(d) where the incident magnon frequency is $\omega_{i}/2\pi=34$ GHz and the bound magnon frequency is $\omega_{b}/2\pi=6$ GHz.}
\end{figure}

\begin{figure}[ptbh]
\begin{centering}
\includegraphics[width=0.48\textwidth]{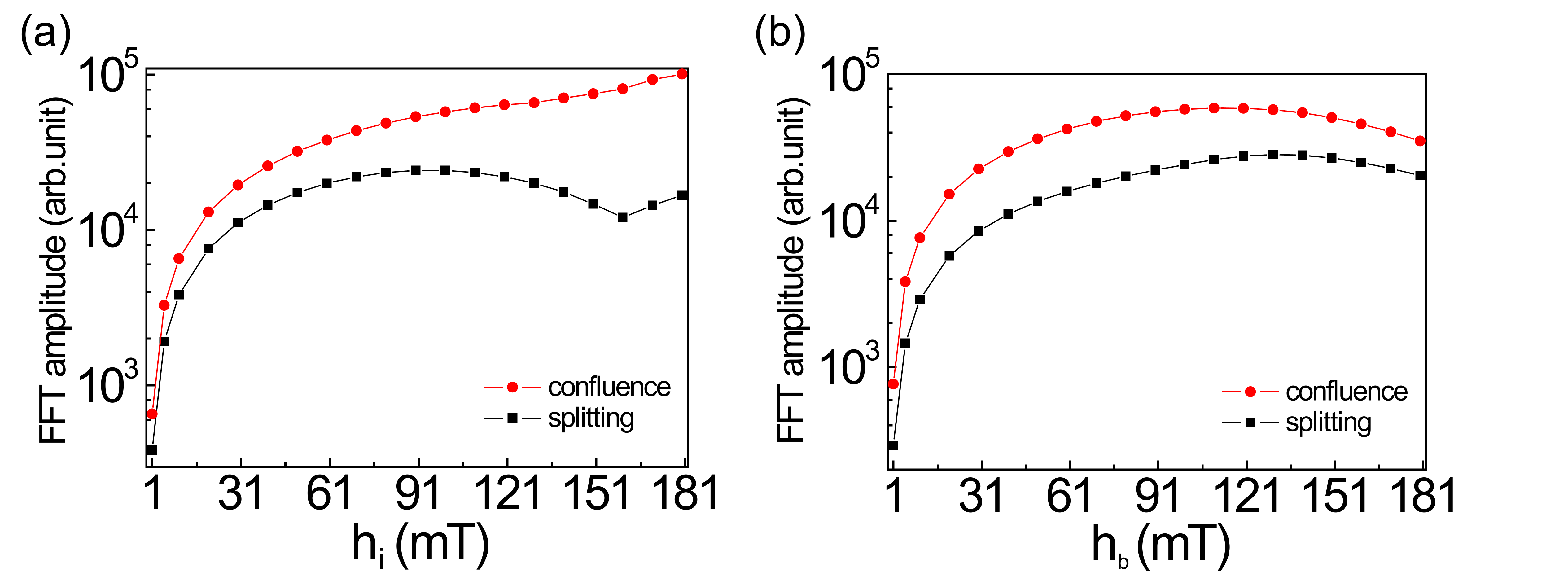}
\par\end{centering}
\caption{Amplitude of spatial FFT as a function of (a) $h_{i}$ by keeping $h_{b}=100$ mT, and (b) $h_{b}$ with fixed $h_{i}=100$ mT, for the confluence peak at 40 GHz (red circles) and the splitting peak at 28 GHz (black squares) observed in Fig. 8. Other parameters are $\omega_{i}/2\pi=34$ GHz and $\omega_{b}/2\pi=6$ GHz.}
\end{figure}

In the above calculations, we have studied the three-magnon events under fixed microwave fields, while it is not clear how the nonlinear processes are modulated by the strength of driving fields. To this end, we systematically calculate the spatial FFT amplitude as functions of $h_{i}$ and $h_{b}$, under fixed input frequencies $\omega_{i}/2\pi=$ 34 GHz and $\omega_{b}/2\pi=$ 6 GHz. Figure 9(a) shows the $h_{i}$-dependence of the spatial FFT amplitudes by fixing $h_{b}=100$ mT. We find that the confluence amplitude increases with increasing field. The amplitude of splitting process shows a similar field dependence when the field is below 100 mT. These results are consistent with our analysis, particularly in the very low field region where we have the FFT amplitude $\propto$$\sqrt{I}$ and the field strength $\propto$$\sqrt{n}$, with $I$ the intensity of the nonlinear process and $n$ the magnon number discussed in Secs. IIA and IIB. The splitting amplitude then decreases with a dip at 160 mT and increases again with respect to $h_{i}$, which may involve higher-order nonlinear processes. The $h_{b}$-dependence of the spatial FFT amplitudes is plotted in Fig. 9(b), in which we keep $h_{i}=100$ mT. It shows that the amplitude of the confluence (stimulated splitting) process monotonically increases with the driving field $h_{b}$ until 110 mT (130 mT) and decreases subsequently, which indeed supports the view that the presence of bounded mode stimulates a significant enhancement of the three-magnon splittings.

Before concluding this article, we discuss the effect from the magnonic spin transfer torque \cite{Yan} which was not addressed in the above analysis. In the micromagnetic simulations, we indeed observed a spin-wave driven domain-wall propagation to the left domain (not shown), with a velocity V$_{\text{DW}}\approx0.84-1.59$ m s$^{-1}$ for all incident frequencies considered in the numerical calculations. This finite domain wall velocity could result in a violation of the energy conservation used in Eqs. (\ref{Confluence}) and (\ref{ConservationSp}), with a value $q$V$_{\text{DW}}\approx0.04-0.1$ GHz smaller than the frequency resolution of FFT.
\section{Conclusion}
To summarize, we theoretically address the interaction between propagating spin-wave modes and localized modes in inhomogeneous magnetization textures, and propose a scheme to eavesdrop on the spin-wave spectrum confined in the domain-wall nanochannel via nonlinear three-magnon processes. The three-magnon confluence process is routine, while the three-magnon splitting is highly nontrivial. We uncover a stimulated three-magnon splitting effect, assisted by the bounded magnon moving in the wall channel. Our theoretical analysis shows that, once knowing the information of injection wave in one magnetic domain and the emerging modes in the opposite domain, we are able to uniquely infer the spectrum of the spin-wave in the nanochannel formed by the domain wall. Micromagnetic simulations agree excellently with analytical formulas. Our results expose an information security issue to the magnonics community by demonstrating a novel non-local method to detect the channelled spin waves.

\section*{ACKNOWLEDGMENT}

We thank H. Yang, C. Wang and X.S. Wang for useful discussions. This work is supported by the National
Natural Science Foundation of China (Grants No. 11604041 and 11704060), the
National Key Research Development Program under Contract No. 2016YFA0300801,
and the National Thousand-Young-Talent Program of China. X.R. Wang is supported by National Natural Science Foundation of China (Grant No. 11374249) and Hong Kong RGC (Grants No. 16301115 and 16301816).\\

\section*{APPENDIX}

We derive the magnon-mgnon interaction in inhomogeneous magnetization textures by starting with the following Hamiltonian
\begin{equation}\label{Energy}
 H=\int d\textbf{r} \bigg\{\frac{A}{M^{2}_{s}} {(\nabla \textbf{M})}^{2}-\frac{D}{M^{2}_{s}} (\textbf{M}\cdot \textbf{n})^2\bigg\},
\end{equation} where $\textbf{M}$ is the magnetization, $A$ is the exchange constant, $D$ is the anisotropy constant, and $\textbf{n}$ is the unit vector along the anisotropy axis (the $z$ axis). \begin{figure}[ptbh]
\begin{centering}
\includegraphics[width=0.3\textwidth]{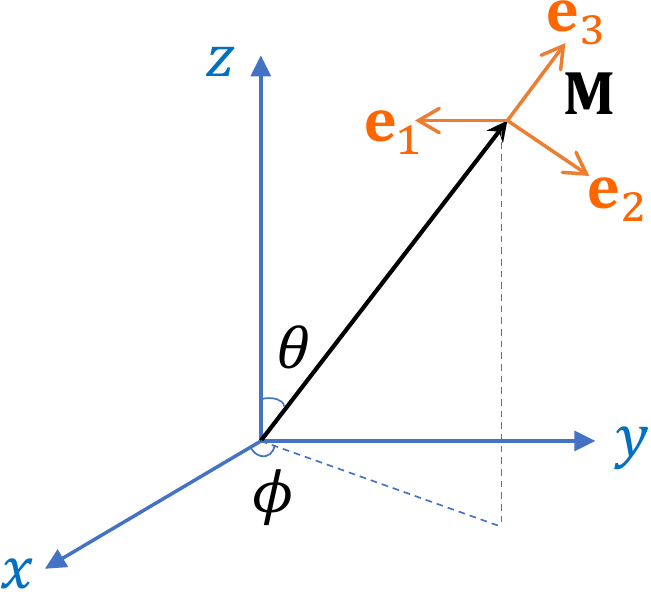}
\par\end{centering}
\caption{Cartesian coordinates in the lab and local frames.}%
\end{figure}We consider small oscillations of the magnetization against the background of a classical spin texture. To this end, we represent $\textbf{M}$ in the form $\textbf{M}_{0}(\textbf{r},t)+\textbf{s}(\textbf{r},t)$, where $\textbf{M}_{0}(\textbf{r},t)$ is the background magnetization distribution, and $\textbf{s}$ corresponds to the small oscillations of the magnetization against $\textbf{M}_{0}$. For simplicity, we do not consider the case of a time-dependent magnetization texture. So, $\textbf{M}_{0}(\textbf{r},t)=\textbf{M}_{0}(\textbf{r})$. It is convenient to introduce a new coordinate system (shown in Fig. 10) in which the axis of quantization $\textbf{e}_3$ for $\textbf{s}$ coincides with the equilibrium direction $\textbf{M}_{0}(\textbf{r})$:

\begin{equation}\label{Transformation}
  \left(
           \begin{array}{c}
             \textbf{e}_1 \\
             \textbf{e}_2 \\
             \textbf{e}_3 \\
           \end{array}
         \right)=\left(
    \begin{array}{ccc}
      \sin\phi & -\cos\phi & 0 \\
      \cos\theta \cos\phi & \cos\theta \sin\phi & -\sin\theta \\
       \sin\theta \cos\phi & \sin\theta \sin\phi & \cos\theta \\
    \end{array}
  \right)\left(
           \begin{array}{c}
             \textbf{e}_x \\
             \textbf{e}_y \\
             \textbf{e}_z \\
           \end{array}
         \right),
\end{equation}

In this system $M_{03}=M_{0},M_{01}=M_{02}=0$; the nonuniform magnetization distribution corresponding to the domain wall is described by the angles $\theta(\textbf{r})$ and $\phi(\textbf{r})$. We thus have
\begin{equation}\label{Mi}
  \begin{aligned}
  M_{x}=&M_{1}\sin\phi+M_{2}\cos\theta\cos\phi+M_{3}\sin\theta\cos\phi,\\
  M_{y}=&-M_{1}\cos\phi+M_{2}\cos\theta\sin\phi+M_{3}\sin\theta\sin\phi,\\
  M_{z}=&-M_{2}\sin\theta+M_{3}\cos\theta,\\
  \end{aligned}
\end{equation}
where $M_{1}=s_{1}, M_{2}=s_{2}, \text{     and     } M_{3}=M_{0}+s_{3}$.

Then, we obtain the expression of the exchange energy
\begin{widetext}
\begin{equation}\label{Exchange}
\begin{aligned}
  \frac{A}{M^{2}_{s}}(\nabla \textbf{M})^{2}=&\frac{A}{M^{2}_{s}} [(\nabla M_{1})^{2}+(\nabla M_{2})^{2}+(\nabla M_{3})^{2}]\\
  +&\frac{A}{M^{2}_{s}}(\nabla\theta)^{2}[(M_{2})^{2}+(M_{3})^{2}]
  +\frac{A}{M^{2}_{s}}(\nabla\phi)^{2}[(M_{1})^{2}+(M_{2}\cos\theta+M_{3}\sin\theta)^{2}]\\
  +&\frac{2A}{M^{2}_{s}} (\nabla\theta)(\nabla\phi)M_{1}(M_{3}\cos\theta-M_{2}\sin\theta)\\
  +&\frac{2A}{M^{2}_{s}} (\nabla\phi)[\sin\theta (M_{1}\nabla M_{3}-M_{3}\nabla M_{1})+\cos\theta (M_{1}\nabla M_{2}-M_{2}\nabla M_{1})]\\
  +&\frac{2A}{M^{2}_{s}} (\nabla\theta)(M_{3}\nabla M_{2}-M_{2}\nabla M_{3}),\\
\end{aligned}
\end{equation}
the anisotropy energy
\begin{equation}\label{Anisotropy}
  -\frac{D}{M^{2}_{s}} (\textbf{M}\cdot \textbf{n})^{2}=-\frac{D}{M^{2}_{s}}(M_{3}\cos\theta-M_{2}\sin\theta)^{2},
\end{equation}
and finally the total energy
\begin{equation}\label{Energy2}
\begin{aligned}
 H=&\int d\textbf{r} \bigg\{\frac{A}{M^{2}_{s}} {(\nabla \textbf{M})}^{2}-\frac{D}{M^{2}_{s}} (\textbf{M}\cdot \textbf{n})^{2}\bigg\}\\
 =&\int d\textbf{r} \bigg\{\sum_{i}\frac{A}{M^{2}_{s}} (\nabla M_{i})^{2}+\frac{A}{M^{2}_{s}}(\nabla\theta)^{2}[(M_{2})^{2}+(M_{3})^{2}]+\frac{A}{M^{2}_{s}}(\nabla\phi)^{2}[(M_{1})^{2}+(M_{2}\cos\theta+M_{3}\sin\theta)^{2}]\\
 &+\frac{2A}{M^{2}_{s}} (\nabla\theta)(\nabla\phi)M_{1}(M_{3}\cos\theta-M_{2}\sin\theta)+\frac{2A}{M^{2}_{s}} \nabla\theta (M_{3}\nabla M_{2}-M_{2}\nabla M_{3})\\
 &+\frac{2A}{M^{2}_{s}} \nabla\phi [\sin\theta (M_{1}\nabla M_{3}-M_{3}\nabla M_{1})+\cos\theta (M_{1}\nabla M_{2}-M_{2}\nabla M_{1})]-\frac{D}{M^{2}_{s}}(M_{3}\cos\theta-M_{2}\sin\theta)^{2}\bigg\}.
 \end{aligned}
\end{equation}
\end{widetext}
We shall express the components of $\textbf{s}(\textbf{r},t)$ in the rotated coordinates in terms of the Holstein-Primakoff operators $a(\textbf{r})$ and $a^{+}(\textbf{r})$:
\begin{equation}\label{HP}
   \begin{aligned}
   s^{+}=&s_{1}+i s_{2}=2\sqrt{\mu_{B}M_{s}}(1-\frac{\mu_{B}a^{+}a}{M_{s}})^{\frac{1}{2}}a,\\
   s^{-}=&s_{1}-i s_{2}=2\sqrt{\mu_{B}M_{s}}a^{+}(1-\frac{\mu_{B}a^{+}a}{M_{s}})^{\frac{1}{2}},\\
   s_{3}=&-2\mu_{B}a^{+}a,\\
   \end{aligned}
\end{equation}
where $\mu_{B}$ is the Bohr magneton. The operators $a(\textbf{r})$ and $a^{+}(\textbf{r})$ satisfy the Bose commutation relations
\begin{equation}\label{Commutation}
  [a(\textbf{r}),a^{+}(\textbf{r}')]=\delta(\textbf{r}-\textbf{r}'),
\end{equation}
and are the annihilation and creation operators of spin waves. We then have
\begin{equation}\label{M123r}
  \begin{aligned}
  M_{1}=&\mu_{B}\sqrt{2S}[(1-\frac{a^{+}a}{2S})^{\frac{1}{2}}a+a^{+}(1-\frac{a^{+}a}{2S})^{\frac{1}{2}}],\\
  M_{2}=&-i\mu_{B}\sqrt{2S}[(1-\frac{a^{+}a}{2S})^{\frac{1}{2}}a-a^{+}(1-\frac{a^{+}a}{2S})^{\frac{1}{2}}],\\
  M_{3}=&2\mu_{B}(S-a^{+}a),\\
  \end{aligned}
\end{equation}
with $S=M_{s}/(2\mu_{B})$ the spin of an atom. Substituting Eqs. (\ref{M123r}) into the total energy (\ref{Energy2}), we obtain a formal expansion of the resulting bosonic Hamiltonian with a small parameter $1/S$:
\begin{widetext}
\begin{equation}\label{Expansion}
  H=S^2E_{\text{class}}+SH^{(2)}+\sqrt{S}H^{(3)}_{\text{int}}+S^{0}H^{(4)}_{\text{int}}+\cdots.
\end{equation}
The first term $E_{\text{class}}$ corresponds to the classical energy of the ferromagnet,
\begin{equation}\label{Classical}
  S^2E_{\text{class}}=\int d\textbf{r} \bigg\{A[(\nabla\theta)^{2}+\sin^{2}\theta(\nabla\phi)^{2}]+D\sin^{2}\theta\bigg\}.
\end{equation}
The second term, $H^{(2)}$, in (\ref{Expansion}) is quadratic in boson operators $a(\textbf{r})$ and $a^{+}(\textbf{r})$, and describes the linear spin-wave theory on top of inhomogeneous magnetization textures. The expression of $H^{(2)}$ is presented in the following
\begin{equation}\label{Secondorder}
  \begin{aligned}
  SH^{(2)}=&2S^{-1}\int d\textbf{r}\bigg \{A(\nabla a^{+}\nabla a)+a^{+}a[D-\frac{A}{2}(\nabla \theta)^{2}-\frac{3D}{2}\sin^{2}\theta+\frac{A}{2}(\nabla \phi)^{2}(1+\cos^{2}\theta)] \\
           &+\frac{1}{4}(a^{+}a^{+}+aa)[(A(\nabla \phi)^{2}+D)\sin^{2}\theta-A(\nabla \theta)^{2}]+\frac{iA}{2}[(-a^{+}a^{+}+aa)\nabla \theta\nabla \phi+(-a^{+}\nabla a+a\nabla a^{+})\cos\theta\nabla\phi]\bigg \},
  \end{aligned}
\end{equation}
\end{widetext}
which is applicable to arbitrary inhomogeneous magnetization textures.
The forms of $H^{(3)}_{\text{int}}$ and $H^{(4)}_{\text{int}}$ are very complicated for general magnetization textures. We thus consider a N\'{e}el domain wall structure with the magnetization profile described by Eq. (\ref{Walker}) in the main text, and obtain
\begin{widetext}
\begin{equation}\label{DWHint}
  \begin{aligned}
     SH^{(2)}=&2S^{-1}D\int d\textbf{r}\bigg\{w^{2}(\nabla a^{+})(\nabla a)+[1-\frac{2}{\cosh^{2}(\frac{y-Y}{w})}]a^{+}a\bigg\},\\
     \sqrt{S}H^{(3)}_{\text{int}}=&i2\sqrt{2}S^{-3/2}D\int d\textbf{r}\bigg\{wa^{+}a\frac{d}{dy}[\frac{a^{+}-a}{\cosh(\frac{y-Y}{w})}]\bigg\},\\
     S^{0}H^{(4)}_{\text{int}}=&S^{-2}D\int d\textbf{r}a^{+}a\bigg\{w^{2}(\nabla a^{+})(\nabla a)+[1-\frac{2}{\cosh^{2}(\frac{y-Y}{w})}]a^{+}a\bigg\},\\
     \vdots
  \end{aligned}
\end{equation}
\end{widetext}

\end{document}